\documentstyle[mnextra,psfig]{mn}

\seceqnum
\nobrackets

\title[Measuring $\Omega_0$ from the Entropy Evolution of Clusters]
      {Measuring $\Omega_0$ from the Entropy Evolution of Clusters}

\author[Scott T. Kay and Richard G. Bower]
       {Scott T. Kay$^{1,2}$ and Richard G. Bower$^{1,3}$  \\
	$^1$ Physics Department, 
	University of Durham,
	Science Laboratories,
	South Rd, Durham DH1 3LE\\
	$^2$ Scott.Kay@durham.ac.uk\\
	$^3$ R.G.Bower@durham.ac.uk}

\date{\today}

\def\ROSAT{{\it ROSAT }}
\def\ASCA{{\it ASCA }}
\def\AXAF{{\it AXAF }}
\def\XMM{{\it XMM }}
\def\Einstein{{\it Einstein }}

\def\EMSS{{\it EMSS }}

\def\SHARC{{\rm SHARC }}
\def\RDCS{{\rm RDCS }}
\def\WARPS{{\rm WARPS }}

\def\MNRAS{MNRAS}
\def\ApJ{ApJ}
\def\ApJS{ApJS}
\def\AA{A\&A}

\def\Science{Science}
\def\AN{AN}

\def\hmsun{\,\hbox{h}^{-1}\,\hbox{M}_{\odot}}
\def\hMpc{\,\hbox{h}^{-1}\,\hbox{Mpc}}
\def\Mpc{\,\hbox{Mpc}}
\def\ergs{\,\hbox{erg}\,\hbox{s}^{-1}}

\def\keV{\,\hbox{keV}}

\def\kpc{\,\hbox{kpc}}
\def\hkmsMpc{\,\hbox{h}\,\hbox{km}\,\hbox{s}^{-1}\,\hbox{Mpc}^{-1}}
\def\kmsMpc{\,\hbox{km}\,\hbox{s}^{-1}\,\hbox{Mpc}^{-1}}

\def\h50{\,\hbox{h}_{50}}

\def\ie{{i.e. \,}}
\def\eg{{e.g. \,}}
\def\cf{{c.f. \,}}

\def\Nbody{$N$-body }
\def\smin{\,\hbox{s}_{\rm min}}
\def\betamodel{\beta-\hbox{model}}
\def\rvir{r_{\rm vir}}
\def\rhovir{\rho_{\rm vir}}

\begin{document}
\maketitle

\begin{abstract}
In this paper, we have extended the entropy--driven model of cluster evolution 
developed by Bower (\shortcite{Bower97}) in order to be able to predict the 
evolution of galaxy clusters for a range of cosmological scenarios. We have
applied this model to recent measurements of the evolution of the $L_{\rm x}-T$
normalisation and X-ray luminosity function in order to place constraints on
cosmological parameters. We find that these measurements alone do not select
a particular cosmological frame--work. An additional constraint is required
on the effective slope of the power spectrum to break the
degeneracy that exists between this and the background cosmology.
We have therefore included a theoretical calculation of the $\Omega_0$ 
dependence on the power spectrum, based on the cold dark matter paradigm,
which infers $\Omega_0<0.55$ $(0.1 < \Omega_0 < 0.7$ for $\Omega_0+\Lambda_0=1$),
at the $95\%$ confidence level. Alternatively, an independent measurement of the 
slope of the power spectrum from galaxy clustering requires $\Omega_0 < 0.6$ 
($\Omega_0 < 0.65$ for $\Omega_0+\Lambda_0=1$), again to $95\%$ confidence. The 
rate of entropy evolution is insensitive to the values of $\Omega_0$ considered, 
although is sensitive to changes in the distribution of the intracluster medium.  
\end{abstract}


\section{Introduction}
Clusters of galaxies are the largest virialised mass concentrations 
in the present--day Universe. Thus, evolutionary studies offer a unique 
method for directly determining the rate at which such structures
grow in mass. This is influenced by the competing effects of the
steepness of the density fluctuation power spectrum, characterised by
the effective slope, $n$, and the current values of the cosmological
parameters, $\Omega_0$ and $\Lambda_0$
\footnote{$\Lambda_0 \equiv \Lambda / 3 H_0^{2}$, with the Hubble constant taking the
value $H_0=100 \hkmsMpc$}.
In this paper, we explore the possibility that the value of $\Omega_0$, or the
combination, $\Omega_0+\Lambda_0=1$ (which produces a flat geometry), can be 
robustly determined from an approach based solely on the X--ray evolution of galaxy clusters.
In order to be successful however, such an approach needs to overcome
two substantial obstacles. Firstly, the correspondence between
cluster mass and X--ray luminosity is not direct, being sensitive to
the way in which the gas fills the cluster's gravitational potential well.
Secondly, the degeneracy between the cosmological parameters and the effective
slope of the power spectrum on cluster scales, $n$, must be broken before a unique value
can be selected. 
The first problem can be tackled by measuring 
the rate of luminosity evolution and calibrating the efficiency of
X--ray emission by some other means of mass estimation, such as
the luminosity--temperature relation (\eg \cite{MS97})
or gravitational lensing effects (\cite{Smail97}, \cite{BS97}).
The second problem can be approached in several ways. The most
straightforward is to adopt a fluctuation spectrum on the grounds of a physical
hypothesis, for example the cold dark matter (CDM) model (\eg \cite{BBKS}).
Alternatively, an empirical measurement of the power spectrum could be used, for
example using the large--scale distribution of rich clusters from redshift surveys
(\eg \cite{TED97}) or by derivation 
from the shape of the cluster temperature distribution function 
(\eg \cite{Ouk97}). Perhaps the most appealing approach would be to
use X--ray observations spanning a wide range in redshift and
luminosity to separate out the models purely on the basis of 
their observed evolution.

Numerous approaches to the determination of $\Omega_0$ have already been 
presented in the literature (\eg see \cite{Dek96}, for a recent review), 
which range from direct arguments based on the 
peculiar motions of galaxies in the local Universe, through indirect
methods using the baryon fraction in clusters of galaxies,
to measurements of the acoustic peak in the microwave background
spectrum. Each of these approaches has their own
strengths and measures a different aspect of the overall cosmological
model. A value based on the rate of cluster mass growth is appealing 
since it measures $\Omega_0$ on the basis of its large--scale effect
over a modest factor of the Universe's expansion. Convergence of all these 
methods will act as confirmation that our global cosmological 
picture is valid and that no crucial additional physical processes
have been omitted.

Discussion of the implications of cluster evolution is also timely
given the growing area of sky that has now been exploited in X--ray 
surveys. These range in strategy from wide--area projects based
on the \ROSAT all--sky survey (\cite{E97}, \cite{DeG97}),
\EMSS (\cite{H92}, \cite{GL94}), and the \ROSAT North
Ecliptic Pole survey (\cite{HENRYNEP99}), to smaller solid--angle surveys 
based on serendipitous sources identified in deep \ROSAT fields 
(\eg \cite{Ros98},  \cite{Scharf97}, \cite{Coll97}).
Some constraints are also available 
from very deep survey fields (\cite{H98}, \cite{McH98}, \cite{Bower96}) although the area
covered by these fields is currently
very small. In addition, reliable temperature data is also becoming
available for clusters spanning a range of redshifts and luminosities, via the \ASCA
satellite (\eg \cite{Tsuru96}, \cite{MS97}, \cite{Mark98}). 
These recent advances have motivated our study, but there is also a clear
need to identify the places where future observations, based
for example on the forthcoming \AXAF and \XMM missions
are best targeted. One crucial question is whether more progress is to be
made by going to lower flux levels, or more uniform surveys 
covering a wider area of sky.

Our work is related to that of several others in recent literature, 
representing a range of possible approaches to the problem
of cluster evolution. This paper uses a phenomenological model
that separates out the evolution of clusters into factors depending on both the
evolution of the mass spectrum and processes resulting in heating and cooling
of the intracluster gas. 
The foundations of the model were discussed extensively in \cite{Bower97} 
(hereafter Paper~I), providing both a physical basis and interpretation.
It allows us to minimise additional theoretical input into the 
calculation, by using simple scaling relations to translate the properties
of the cluster sample at low redshifts into their equivalents at
earlier epochs. The approaches used by Mathiesen \& Evrard 
(\shortcite{Math98}), Reichart et al.\ (\shortcite{Reich}) and Blanchard \& Bartlett (\shortcite{BB98}) are related, but use an empirical model
for the X--ray luminosity calibration and evolution, combined
with the Press--Schechter method (\cite{PS74}) for the 
distribution of cluster masses. The approach of Kitayama \& Suto (\shortcite{KS97}) is more
different in the sense that they achieve a match to the luminosity function
data through varying the epoch at which the clusters must form. In the 
case of very low values of $\Omega_0$, the distinction between the 
epoch at which a cluster is observed and that at which it is 
formed becomes important. We have therefore developed the model
from Paper~I to incorporate this effect. Finally, a method
of direct deconvolution has been outlined by Henry (\shortcite{HENRY}, see also 
\cite{ECFH}) using recent \ASCA temperature data.
A wide variety of approaches to this topic is clearly desirable in order 
to indicate the robustness of the underlying principles. We have therefore 
included in our discussion, a comparison between our work and others
and we outline the areas of uncertainty that can be considerably 
improved from further observations.

The layout of this paper is as follows: 
in \S2, we outline the 
model on which this paper is based and show how it can be readily
extended to incorporate evolution in both open ($\Omega_0<1$,\ $\Lambda_0=0$) and 
flat ($\Omega_0<1$,\ $\Omega_0+\Lambda_0=1$), sub--critical density universes. We have
also included in this section, our approach to incorporate the effects of the
cluster formation epoch.
\S3 summarises the constraints on the model parameters using currently
available X--ray cluster data for a range of cosmologies and we investigate
the limits that can, at present, be set by this approach. We also focus on the
inherent degeneracy that exists between $\Omega_0$ and $n$, and investigate 
whether we can empirically distinguish between different cosmological models,
using the methods mentioned earlier. The possibility of measuring evolution
of the cluster core radius to place further constraints on the model is
also discussed.
In \S4 we summarise our results and investigate the robustness of assumptions 
made in this model. We also compare our results
with the values that have been obtained by other authors, and explore the 
differences between the proposed models leading us to consider the overall 
accuracy of the method. With this in mind, we identify key strategies for future
X--ray surveys. Finally, in section \S5 we reiterate our conclusions.

\section{Modelling the X--ray Evolution in Different Cosmological Scenarios}

\subsection{X--ray Emission and the Cluster Core}
It is now well established that X--ray emission from the intracluster medium
is dominated by collisional processes. For hot clusters, the most
important of these is thermal bremsstrahlung.  The emissivity scales 
as $\rho^2 T^{\alpha}$, where we use
$\alpha = 0.4\,[\pm 0.1]$ for bolometric/wide--band detectors and $\alpha = 0\,[\pm 0.1]$ for
low energy band--passes, although we find that the results are not affected when
making changes to these values within the limits quoted in square brackets.
 Detections of galaxy clusters in this region
of the electromagnetic spectrum are sensitive to the way in 
which the intracluster gas is distributed. Surface brightness distributions are 
generally fitted using the following (so--called $\betamodel$) density profile
\begin{equation}
 \rho(r)=\rho_c\left[1 + \left({r \over r_c}\right)^2\right]^{-\frac{3}{2}\beta},
\label{eqn:denprof}
\end{equation}
with $r_c$ defining the effective core--size and $\beta$ the rate at which the
density falls off with radius. We adopt a constant value of $\beta=2/3$
for the main results in this paper. This is appropriate for a non--singular isothermal
distribution and is in agreement with the observational average
(\eg \cite{JF84}). The effect of departures from this
asymptotic slope is discussed in \S4.1.
The emission is thus characterised by a flattening at small scales
(typically $r_c \sim 100 \kpc)$, departing from the distribution expected
if the gas followed the underlying dark matter.
A plausible physical interpretation was given by Evrard \& Henry (\shortcite{EH91},
hereafter EH,
see also \cite{K91}), hypothesising that the intracluster gas was pre--heated
prior to the cluster's formation and has retained the entropy acquired
from this pre--collapse phase. Assuming that any dissipation was negligible,
this provides an entropy {\it floor}, $\smin$
\footnote{we use the definition of specific entropy, 
$s \equiv c_v{\rm ln}(T \rho^{1-\gamma})$ ; $\gamma$ is assumed
to be $5/3$, appropriate for a non--relativistic ideal gas},
forcing the gas to build up a mass distribution that reflects this constraint.
For an isothermal distribution, $\smin$ directly corresponds to the 
core density, $\rho_c$. In paper~I, this idea was developed
further by allowing $s_{\rm min}$ to evolve, using the parameterization
\begin{equation}
    \label{eq:sminz}
  s_{\rm min} = s_{\rm min}(z=0)+c_v \, \epsilon \, \rm{ln}(1+z),
\end{equation}
where $\epsilon$ determines the rate of core entropy evolution, which 
could be dominated by, for example, the gas being shock--heated as a result
of merging (implying negative values) or conversely, radiative cooling 
(positive values). The value $\epsilon=0$ corresponds to no net evolution
of the core entropy. 

\subsection{Evolution in an $\Omega=1$ Universe}
The case of evolution in a critical density universe was discussed
extensively in paper~I. We outline the details again here before going
on to discuss evolution in the general cosmological context.
The X--ray luminosity of an individual cluster is calculated by
integrating over the virialised region 
\begin{equation}
L_{\rm x} \propto \int\limits_{0}^{\rvir} r^{2} \, \rho^{2} \, T^{\alpha} dr. 
\end{equation}
We use Eq.~\ref{eqn:denprof} to describe the density profile, and assume
that the temperature profile of the gas distribution can be described in a
similar way, with the characteristic temperature in proportion to the 
virial temperature of the system. In practice,
the integral is dominated by the contribution from within
a few core radii, and thus the scaling properties of this integral depend
weakly on the assumed density profile. Furthermore, departures from
the standard profile can be accommodated by redefining the core radius
of the system. As in EH, extracting the scaling properties from the integral
produces the following result:
\begin{equation}
\label{eq:lxscale}
L_{\rm x} \propto \rho_c^2 \, r_c^3 \, T^{\alpha}.
\end{equation}

Using Eq.~\ref{eq:sminz} to parametrise the evolution of the core entropy, 
the core density evolves as 
\begin{equation}
  \rho_c \propto T^{\frac {3}{2}}\ (1+z)^{-\frac{3 \epsilon}{2}}.
\end{equation}
The core density can also be related to the virial radius, virial density
and the core radius, using the asymptotic profile
of the gas distribution, 
$
 \rho_c \sim \rhovir \left( \frac {\rvir}{r_c} \right)^{3\beta}
$. Note that the effectiveness of a certain value of $\epsilon$ is dependent on
the $\beta$--profile adopted. Coupled with the definition of the virial 
radius,
$\rvir \propto (M \rhovir^{-1})^{\frac{1}{3}}$ (where $M$ is the virial 
mass of the cluster) this leads to a relation for the characteristic 
cluster core radius
\begin{equation}
 r_c \propto M^{\frac{1}{3}} \ T^{-\frac{1}{2\beta}} \
 (1+z)^{-1+\frac{1}{\beta}(1+\frac{\epsilon}{2})}.
\end{equation}
We combine these results with the scaling relations for mass
($M \propto (1+z)^{-6 \over n+3}$) and virial temperature 
($T \propto M^{\frac{2}{3}} \ (1+z)$) to obtain the redshift dependence
of the cluster luminosity, 
\begin{equation}
L_x \propto (1+z)^{3(\frac{1}{\beta}(1+\frac{\epsilon}{2})-\epsilon-\frac{3}{2})} \ T^{\alpha+
\frac{3}{2}(3-\frac{1}{\beta})}
\end{equation}
and number density, $n \propto (1+z)^{6 \over n+3}$.

The above relations are not appropriate for the scaling properties of individual
clusters. The histories of particular objects will differ wildly.
Rather, we use only the Weak Self--Similarity Principle to apply these
scalings to the mean evolution of the population as
a whole (\cf Paper~I). This is strictly correct if the universe has
critical density and a scale free spectrum of density perturbations.
Note, however, that measurements based on a single cluster population
(\eg the slope of the $L_{\rm x}-T$ relation) 
cannot be constrained using this approach since the formation histories
of clusters with greater or smaller masses than the characteristic value may be
different.

\subsection{Extension to $\Omega_0<1$ Cosmologies}
\label{sec:modelx}

For a sub--critical density Universe, several modifications are required.
Firstly, the Weak Self--similarity Principle cannot be rigorously
justified in a low density universe 
since the internal mass structure of the clusters need not
be homologous and will depend on the epoch at which they collapse. 
In \Nbody simulations, however, deviations from homology are small 
(\eg \cite{ENF98}, for the $\Omega_0 = 0.3, \Lambda_0 = 0.7$ scenario);
we will assume that any departure from homology can be incorporated into
the definition of the $\epsilon$ parameter. For the aim of this paper, ie.
comparing the evolutionary properties of clusters in critical and low density 
universes, this approach is adequate. 

The characteristic density of the cluster system will however depend
on the epoch at which the system collapses ($z_f$). For a very low 
density Universe, this may be significantly different from the epoch at
which the cluster is observed.
To relate the dynamical properties of cluster populations to the background properties
set by the cosmology, we use 
the results of the spherical top--hat collapse model (\eg \cite{Lahav91}). 
This predicts that that overdense 
spherically symmetric perturbations depart from the linear regime,
turn around and collapse, forming virialised structures with mean internal
densities given by the formula
$\rho_{\rm vir} = \Delta_{\rm vir}\ \rho_{\rm crit}$, where $\rho_{\rm crit}$
is the density required to close the Universe.
The quantity $\Delta_{\rm vir}$ is a function of the background cosmology and
can be approximated by using the following fitting functions (\cite{ENF98}):
\begin{eqnarray}
\Delta_{\rm vir}&=& \, 178 \ \Omega^{0.30} \qquad (\Lambda_0 = 0)\nonumber\\
		&=& \, 178 \ \Omega^{0.45} \qquad (\Omega_0 + \Lambda_0 = 1).
\end{eqnarray}

Finally, the linear growth factor, $\delta_{+}$, evolves more slowly than $(1+z)^{-1}$
for $\Omega_0<1$ models. The appropriate relations are as follows (\cite{LSSBible})
\begin{eqnarray}
\Omega_0 < 1 :- \nonumber\\
\delta_{+}(x)&=&1+\frac{3}{x}+{3(1+x)^{\frac{1}{2}} \over x^{\frac{3}{2}}}
{\rm ln}[(1+x)^{\frac{1}{2}}-x^{\frac{1}{2}}], \nonumber\\
x&=&{\Omega_0^{-1}-1 \over 1+z}. \nonumber \\
\Omega_0 + \Lambda_0 = 1 :- \nonumber\\
\delta_{+}(y)&=&{(y^3+2)^{\frac{1}{2}} \over y^{\frac{3}{2}}}
\int\limits_{0}^{y}{u \over u^3+2}^{\frac{3}{2}} \ du, \nonumber\\
y&=&{2(\Omega_0^{-1}-1)^{\frac{1}{3}} \over 1+z}. 
\end{eqnarray}
Accounting for these differences produces the following scaling relation
for the X--ray luminosity:
\begin{eqnarray}
\label{eqn:thebigmother}
L_{\rm x} &\propto& T^a \, (1+z)^b \, (1+z_f)^c \, 
                  \left[{\Delta_{\rm vir}(z_f) \over \Omega (z_f)}\right]^d,
\\ \nonumber
a&=&\alpha + \frac{3}{2} \left( 3 - \frac{1}{\beta} \right), \\ \nonumber
b&=&3\epsilon \left( \frac{1}{2\beta}-1 \right), \\ \nonumber
c&=&3\left(\frac{1}{\beta} - \frac{3}{2} \right), \\ \nonumber
d&=&\frac{1}{\beta}-\frac{3}{2}.
\end{eqnarray}
The temperature scaling relation is thus
\begin{equation}
T \propto \delta_{+}(z)^{4 \over n+3} \, (1+z_f) \, 
\left[{\Delta_{\rm vir}(z_f) \over \Omega (z_f)}\right]^{\frac{1}{3}},
\end{equation}
and the number density scales as $ \delta_+^{-{6 \over n+3}}$.

\subsection{The Epoch of Cluster Formation}

\begin{figure}
\centering
\centerline{\psfig{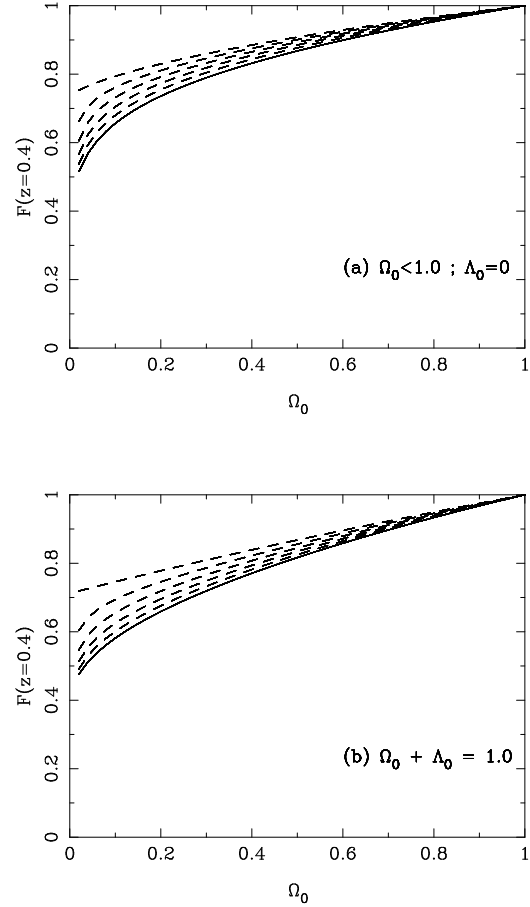}}
\caption{The X--ray luminosity factor that depends on the cluster formation redshift,
 $z_f$, as a function of the background cosmology, for both open ($\Omega_0 < 1$)
and flat ($\Omega_0 + \Lambda_0 = 1$) models, for the case $z=0.4$. 
The solid line illustrates the variation
of F for a mass fraction of 1 and the dashed lines show progressively lower values,
in steps of 0.1, to $f=0.5$ .}
\label{zform}
\end{figure}

Since the linear growth of fluctuations `freezes out' when $z \sim \Omega_0^{-1}-1$
(\cite{LSSBible}), producing the present--day abundance of clusters requires
structure to have formed at progressively earlier epochs for lower density
Universes. Consequently, the epoch of cluster formation, $z_f$ can be significantly
different from the redshift range of an observed sample, leading to inaccuracies
in the scaling properties (\cite{KS96}). We tackle this problem by defining the 
epoch of formation to be such that the cluster has acquired a fraction, $f$, of it's
total mass when observed
\begin{equation}
M(z_f(z)) = f \, M(z).
\end{equation}
Material that has been accumulated between $z_f$ and $z$ is assumed to have
negligible effects on the normalisation of the mass distribution, even though
the total mass and therefore temperature of clusters can change. 

The effect that varying $f$ has on the X--ray luminosity is illustrated in Figure
\ref{zform} for a typical median redshift of current distant cluster samples
($z=0.4$). This is done by plotting the following function
\begin{equation}
F(\Omega_0,\Lambda_0,z)=F_0 \ (1+z_f)^{\frac{11}{4}} 
\left[{\Delta_{\rm vir}(z_f) \over \Omega (z_f)}\right]^{\frac{11}{12}}
\end{equation}
varying with $\Omega_0$, \ie the part of equation \ref{eqn:thebigmother} 
that depends on $z_f$. The constant, $F_0$, normalises the relation to the
corresponding value at the present day ($z_f(z=0)$). Clearly for the
$\Omega=1$ case, the formation epoch scales in direct proportion to the
observed redshift, with the normalisation only depending on the slope of
the power spectrum, $n$ (which also controls the rate at which structure
grows). As the matter density drops below the critical value, we see
a decrease in about a factor of 2 for $\Omega_0=0.1$, although the 
dependence on $f$ is relatively weak (even for $\Omega=0.1$) : reducing $f$
from $1 \rightarrow 0.5$ produces only a 40\% change in $F$, even in this extreme case.
Therefore, we have selected a fiducial value of $f=0.5$ for the results that follow.

\section{Using The Evolution of Clusters to Constrain $\Omega_0$}

The essence of our approach is to scale characteristic properties of a cluster
population from one epoch to another.
We therefore need to use data based on local samples as the basis for 
the scaling transformations described in the last section. We can
then determine the likelihood distribution of parameters by fitting the scaled relations
to the data available at higher redshift. At this stage, we have assumed that our
free parameters are the slope of the power spectrum, $n$, and the entropy evolution
parameter, $\epsilon$. We have chosen to fix the cosmological density at four values
(with and without the cosmological constant) : $\Omega_0 = 0.1,0.3,0.5,1.0$, in
order to clearly illustrate the effects of a varying cosmological background on the
physical evolution of clusters. Below, we discuss the data--sets used to constrain
the set ($n,\epsilon$) and the corresponding results. All data presented have been
compiled with an assumed Hubble constant of $H_{0} = 50 \kmsMpc$.

\subsection{The Luminosity--Temperature Relation}

\begin{table}
\begin{center}
\begin{tabular}{l|l|l|l}
\hline
Parameter	& Best Fit	& $95\%$ (min,max) \\
\hline
$\eta$		& -0.10		& (-0.63,+0.43)\\
$\lambda$	& 0.292		& (0.325,0.260)\\
\hline
\end{tabular}
\caption{Values of the parameters used to fit the (temperature) evolution of
the $L_{\rm x}-T$ relation, assuming a power law dependence on $L_{\rm x}$ with
slope, $\lambda$ and an evolution term between low and high redshift parameterised
as $(1+z)^{\eta}$. Shown are the best--fit values of the slope and evolution parameters
and $95\%$ confidence levels on their dispersion, for the median
redshift of the high--z sample, $\left< z \right>=0.3$, and a value of $q_{0}=0$.}
\label{tb:lt}
\end{center}
\end{table}

We place a constraint on the temperature evolution of clusters by making use of the
$L_{\rm x}-T$ relation, which is at present, adequately described by a fixed power law,
$L_{\rm x}\propto T^{\lambda}$, and fixed intrinsic scatter.
We determine the evolution of the $L_{\rm x}-T$ relation by fitting a maximum
likelihood model to the combined low and high redshift data given by
David et al. (\shortcite{David93}) and Mushotzky \& Scharf (\shortcite{MS97}), compiled mainly
from the \ASCA and \Einstein satellites.
We assume the distribution of temperatures (that lead to considerable scatter in the relation)
are Gaussian distributed, and convert the $90\%$ systematic errors in the temperature measurements
(quoted by the authors) to their equivalent $1\sigma$ values. 
In order to ensure that there was good overlap in luminosity between the high
and low redshift data--sets, we limited the comparison to clusters
with bolometric luminosities greater than $10^{44.5}\ergs$, although
the results are not particularly sensitive to this choice. The
model for the $L_{\rm x}-T$ correlation that we fit includes an adjustable
zero--point, slope and intrinsic scatter, as well as a redshift--dependent
normalisation term that is parameterised as $\Delta\log T_0 = 
\eta\log(1+z)$ (where $T_0$ is a reference temperature). We determine confidence 
limits on the evolution of the normalisation by minimising over the other parameters
and calculating the distribution of $C$ values, where
\begin{equation}
C_{LT} \equiv -2 \sum_{i=1}^{N} \, {\rm ln} \, P \mathit{_i} (\eta),
\label{eqn:C}
\end{equation}
with the subscript $i$ running over each cluster, $P \mathit{_i}$ is the probability 
of measuring the 
cluster with a given temperature, luminosity and normalisation, controlled by $\eta$. We
have assumed that $\Delta C$ is distributed as $\chi^2$ with one free parameter (\cite{Cash}).

This treatment assumes that we are not interested in the slope of the
relation. This is exactly true if the luminosities of the clusters
do not evolve and the evolution of the $L_{\rm x}-T$ relation is due to
the temperature evolution of the clusters alone. In practice, however,
a particular choice of the parameters $(n,\epsilon)$ implies correlated
changes in luminosity and temperature. For simplicity and consistency 
with other work, we have converted
the evolution in both quantities into a change in temperature at fixed
luminosity. This correction involves the slope of the relation, although
we emphasise that this is an artefact of the way in which the 
normalisation is quoted. In principle, each model requires that we
examine the likelihood as a function of a combination of $\eta$ and the
fitted slope, $\lambda$ (which changes with the choice of $(n,\epsilon)$ parameters);
in practice, however, the slope of the relation is sufficiently 
constrained that almost identical results
are obtained if we examine the likelihood as a function of the single
parameter $\eta$ and then simply adopt the best fitting slope in the
model calculation. 

The limiting values of the evolution rate parameter are given in 
Table \ref{tb:lt} along with the corresponding slope of the $L_{\rm x}-T$ relation, for
the median redshift of the high--redshift sample ($\left< z \right> =0.3$) 
and a value of $q_{0}=0$ ($\Omega_0=
\Lambda_0=0$). These values are converted to the appropriate cosmologies when required, 
since the normalisation is affected by the assumed value of $\Omega_0$ through the distance
dependence of the luminosity. In $\Omega_0<1$ cosmologies, the luminosities of clusters
are brighter, pushing the evolution of the temperature normalisation in the positive direction
for higher values of $\Omega_0$. 
The uncertainties in the evolution rate are dominated by the intrinsic 
scatter in the relation. We therefore note that while the evolution of the relation
is statistically well defined, it is sensitive to systematic
error and selection that may have tended to exclude hotter (or colder) clusters.

\subsection{The X--ray Luminosity Function}

\begin{figure}
\centering
\centerline{\psfig{file=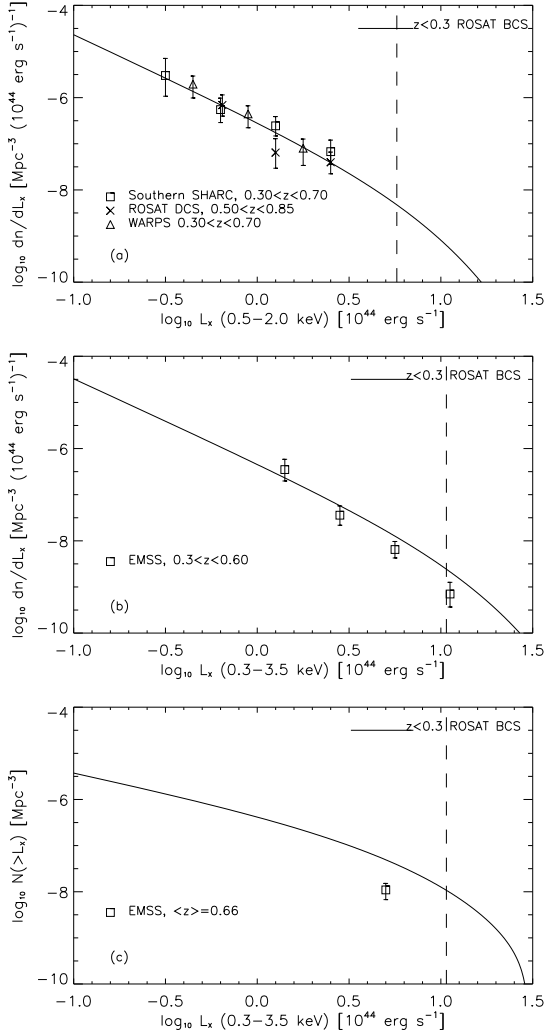,height=14cm}}
\caption{The X--ray luminosity functions used in this paper compiled from data observed 
	using the \ROSAT and \Einstein satellites. The solid line in all three plots
	is the best--fit Schechter function to the \ROSAT BCS $z<0.3$ sample as 
	determined by Ebeling et al. (1997). Plot (a) illustrates the high--redshift
	non--parametric XLF data from \ROSAT surveys, evaluated in the $0.5-2.0 \keV$ 
	pass band.  The square points are from the Southern \SHARC survey (Burke et al. 1997), 
	the crosses from the \RDCS survey (Rosati et al. 1998, private communication) 
	and the triangles from the \WARPS survey (Jones et al. 1998, private communication). 
	Plot (b) illustrates the \EMSS cluster X--ray luminosity function
	from the work of Henry et al. (1992) where plot (c) is the
	cumulative XLF point from the \EMSS distant cluster sample 
	($N(>5 \times 10^{44}\ergs)= 1.1 \pm 0.43 \times 10^{-8} \Mpc^{-3}$; 
	Luppino \& Gioia 1995), both evaluated in the $0.3-3.5 \keV$ band.
	The vertical dashed line marks the position of $L_{\rm x}^{*}$.
	Data is presented assuming $q_{0}=0.5$ ($\Omega=1$).}
\label{fig:combo_xlf}
\end{figure}

The second X--ray observable used in this paper is the cluster X--ray Luminosity function.
For simplicity we adopt a Schechter function parameterization
\begin{equation}
\label{eqn:schechter}
{dn \over dL_{\rm x}} \, = A \, \exp \left( - \frac{L_{\rm x}}{L_{\rm x}^{*}} \right )
                             L_{\rm x}^{-\alpha}. 
\end{equation}

\begin{table}
\begin{center}
\begin{tabular}{l|l|l|l}
\hline
Survey & $N_{bins}$ & $\left< z \right>$ & $N_{clus}$ \\
\hline
\SHARC (\cite{Burke97}) & 4 & 0.44 & 16 \\
\WARPS (\cite{J98})     & 3 & 0.47 & 11 \\
\RDCS  (\cite{Ros98})   & 3 & 0.60 & 14 \\
\EMSS 1(\cite{H92})     & 4 & 0.33 & 23 \\
\EMSS 2(\cite{LG95})    & * & 0.66 &  6 \\
\hline
\end{tabular}
\caption{Details of the distant cluster survey XLF's used for constraining evolution
of the \ROSAT BCS XLF. The \EMSS 2 data is not binned (*) but rather taken as a
point on the cumulative luminosity function.}
\label{tb:survey}
\end{center}
\end{table}

The local XLF used in this paper is the one given by Ebeling et al. (\shortcite{E97}) 
based on the \ROSAT
Brightest Cluster Sample (BCS), which contains clusters with $z<0.3$, selected and
flux--limited at X--ray wavelengths. We adopt their best fit parameters to equation
\ref{eqn:schechter} for the $0.5-2.0 \keV$ band, namely 
$[A=33.2 \, (10^{-8} \Mpc^{-3} \Delta L^{-1}) ; \alpha=1.85 ;
  L_{\rm x}^{*}=5.7 \, (10^{44} \ergs)]$ and
$[A=49.5 \, (10^{-8} \Mpc^{-3} \Delta L^{-1}) ; \alpha=1.82 ; 
  L_{\rm x}^{*}=10.7 \, (10^{44} \ergs)]$
for the $0.3-3.5 \keV$ band. The functions are plotted as solid lines 
in Figure \ref{fig:combo_xlf}.
Since they find no significant evolution ($< 1.8 \sigma$) in their sample,
we assume this to represent the XLF at the present day.

In order to place constraints on the observed evolution of the XLF, we have used the
distant cluster redshift distributions and luminosity functions that are presently
available, summarised in  Table \ref{tb:survey}. Figure \ref{fig:combo_xlf} shows
the positions of the high--redshift data, evaluated for $q_0=0.5 (\Omega=1)$.
Figure \ref{fig:combo_xlf}$(a)$ illustrates
the binned, non--parametric XLF's that have been compiled from \ROSAT data (\SHARC, \WARPS and \RDCS)
, with luminosities evaluated in the $0.5-2.0 \keV$ band. None of these samples show
significant evolution of the cluster XLF, out to typically $z \sim 0.5$. 
Figure \ref{fig:combo_xlf}$(b)$ shows the \EMSS 1
XLF in the $0.3-3.5 \keV$ band. Clearly, this hints at negative evolution 
(\ie a lower space density of clusters of given luminosity at higher redshift), although one
must note that higher luminosity bins usually have greater median redshift values and hence
applying one redshift to the whole sample leads to an overestimation of the evolution.
Finally, Figure \ref{fig:combo_xlf}$(c)$
shows the \EMSS 2 cumulative point, also in the $0.3-3.5 \keV$ band.
Since this point is significantly lower than the BCS XLF, this should provide a tight constraint
on $n$ and $\epsilon$.

We use a likelihood method to compare the data--points with each 
model prediction defined by $(n,\epsilon)$, for the assumed cosmology. 
Ideally, we would constrain our model parameters on the
basis of individual clusters using a global maximum likelihood
approach. Starting from the local XLF, we could assign individual 
likelihood probabilities to
each of the observed clusters (and non--detections) by combining the
model X--ray luminosity function at the appropriate redshift with the
selection function defined by each of the surveys. This approach 
would avoid all problems related to redshift binning of the available
data, and allow cosmological corrections to be consistently applied. 
Unfortunately, the detailed survey selection functions are not generally
available to us, so we must adopt an approximate approach. Since
little evolution is observed, binning the data in redshift is not likely
to result in significant bias. 

The constraints placed on the model parameters $(n,\epsilon)$ from evolution of
the various high--redshift XLF's (relative to the BCS XLF) for a given cosmology
were generated as follows. Firstly, we converted the high--redshift XLF data points
to agree with the assumed cosmology. In $\Omega_0<1$ cosmologies the predicted
space density of clusters decreases because the luminosity limit of the survey is
brighter (scaling as $1/V_{\rm max}$, where $V_{\rm max}$ is the volume limit of the clusters)
, hence the luminosities of the clusters themselves are higher (scaling as $d_L^2$, where $d_L$ 
is the luminosity distance). We scale the space densities assuming that all clusters are at 
the flux limit of the survey, however the correction corresponding to clusters being a factor 
of two brighter is $\sim 1\%$. We then assumed that the number of clusters in the
$i^{th}$ bin ($N_{clus,i}$) were sampled from a Poisson distribution, with the
expected number being a function of the {\it true} values ($n^{T},\epsilon^{T}$). We
calculated the $C$--statistic for the range of model ($n,\epsilon$) such that
$-2.5 \le n \le 1.0$ and $-6 \le \epsilon \le 4$, where $C$ is defined as
\begin{equation}
C_{\rm XLF} = -2 \sum_{i=1}^{N} \, {\rm ln} \, P \mathit{_i} \,(N_{clus,i};E_{i}(n,\epsilon)),
\label{eqn:Cdef}
\end{equation}
$N$ is the number of bins in the high--redshift XLF and
$P_{i}$ is the probability of observing $N_{clus,i}$ clusters given $E_{i}$, the expected number
calculated from the model, at the median redshift of the data point. Confidence
regions were then calculated by differencing $C$ with the minimum value (\ie the most
probable set, ($n_{0},\epsilon_{0}$)), $\Delta C = C-C_{\rm min}(n_{0},\epsilon_{0})$.
If ($n,\epsilon$) were independent parameters, $\Delta C$ would be
distributed as $\chi^{2}$ with two degrees of freedom (\cite{Cash}), but we do not
assume here that this is the case. To circumvent this problem, we generated a large
number of monte--carlo realisations of each XLF data--set, for all the cosmologies studied. 
Each realisation involved generating a set of XLF points at the same
luminosities as the real data--set, drawn from Poisson distributions
with their mean values set to the numbers of clusters predicted by
the best--fit model. We then took the $C$ distribution for each 
realisation and calculated the difference in $C$ values between this
distribution and the one produced by the best--fit model, at the
fixed point ($n_{0},\epsilon_{0}$). This method allows us to 
build up a likelihood distribution of $\Delta C$ values that 
resembles the true distribution. Confidence levels were chosen 
such that they enclosed $68\%$ and $95\%$ of the total number. 

\subsection{Results}
\label{subsec:results}

\begin{figure*}
\centering
\centerline{\psfig{file=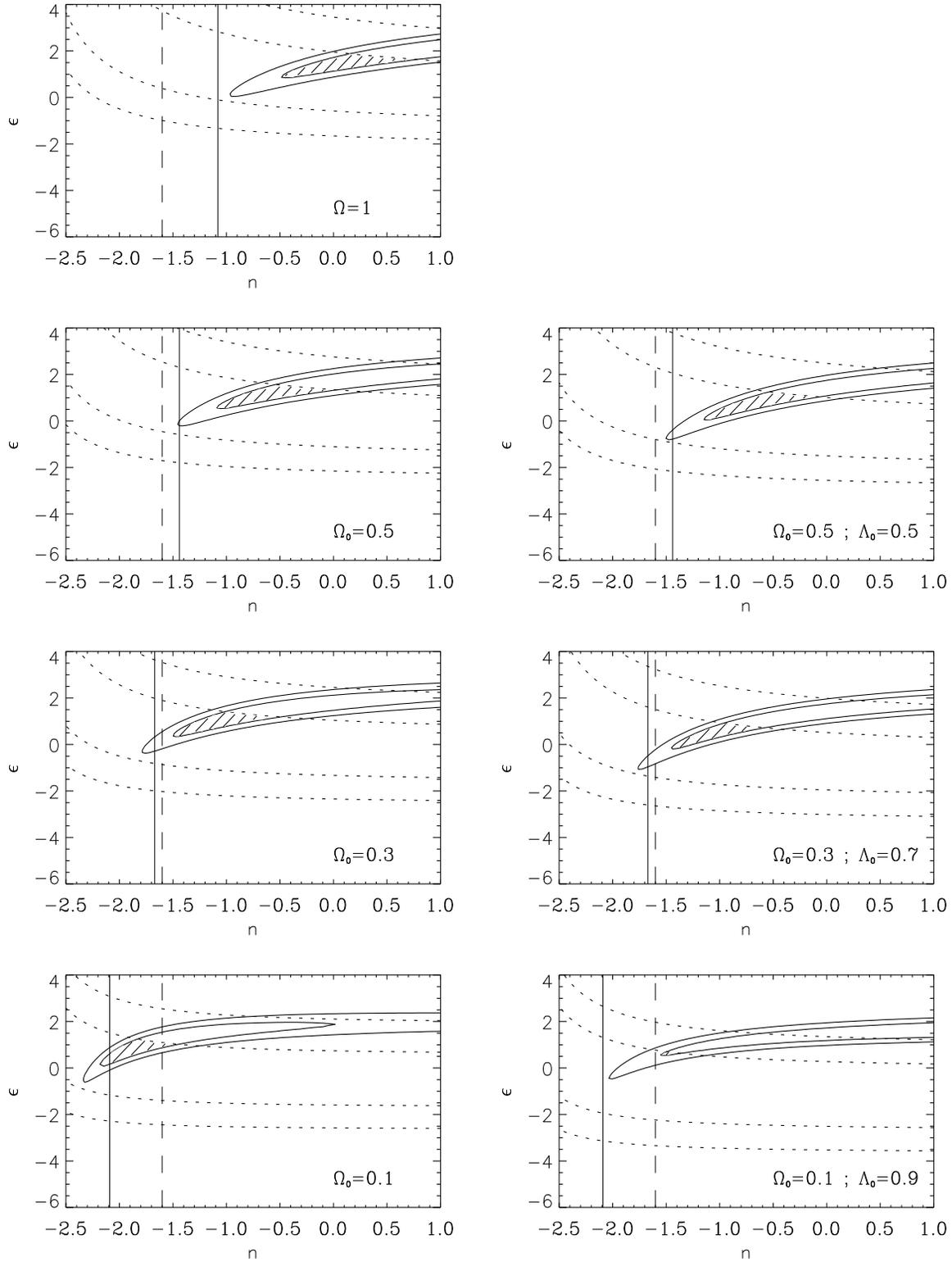,height=21cm,width=16cm}}
\caption{
Constraints on the slope of the power spectrum ($n$) and the entropy evolution
parameter ($\epsilon$) for seven different cosmological scenarios. Solid
contours are the $68\%$ and $95\%$ confidence regions from the evolution of the X--ray luminosity
function, using the combined data from all of the surveys. Dotted 
contours represent the $68\%$ and $95\%$ confidence regions for evolution of
the cluster luminosity--temperature relation. 
Shaded regions are for parameters consistent with the evolution within
$68\%$ limits (of both XLF and $L_{\rm x}-T$). The vertical solid line is the determination
of $n$ from the CDM based theoretical calculation in Section \ref{sec:degen} and 
the vertical dashed line is a measured value of $n$ from the APM rich cluster sample.}
\label{fig:megaplot}
\end{figure*}

For each of the fiducial values of $\Omega_0$ (in both open and flat Universes), 
we determined the likelihood distribution as a function of $n$ and $\epsilon$. 
Initially, we analysed the data--sets independently, however the results show
similar trends, albeit less constrained than the overall likelihood distribution,
generated from combining all of the XLF data. We therefore focus our discussion 
using the combined XLF data--set -- the result of this 
is shown in Figure \ref{fig:megaplot}. Clearly, this illustrates how the constrained
area of parameter space changes with the background cosmology. This
is the result of an interplay between two contributing factors, 
namely the model scaling relations used to fit the high redshift
data and the positions of the data--points themselves, relative to
the local XLF and $L_{\rm x}-T$ relation. 

Constraints from the evolution of the $L_{\rm x}-T$ normalisation
show a weak degeneracy between $n$ and $\epsilon$ for high values
of $\Omega_0$. As $\Omega_0 \rightarrow 0$, the range of $\epsilon$
becomes independent of $n$ (centred on $\epsilon=0$). The range of
$n$ is unconstrained for all values of $\Omega_0$. The reason for 
the degeneracy is straightforward. The data imply no net evolution
of the $L_{\rm x}-T$ normalisation for an empty Universe ($\Omega_0=0$).
Hence, increasing the value of $\Omega_0$ pushes the high redshift
sample to lower luminosities, prompting positive evolution in temperature
for a fixed value of $L_{\rm x}$. In an $\Omega=1$ Universe with no
entropy evolution ($\epsilon=0$) and a power spectrum with a negative slope,
the model over--predicts the cluster luminosities, hence positive entropy evolution 
is required to match the data. As $n$ increases, the over--prediction becomes 
less severe (hence $\epsilon$ decreases). For $\Omega_0<1$ cosmologies, 
the amount of evolution required becomes progressively weaker as $\Omega_0$ 
decreases. Since the growth of structure is also slower, the data becomes more 
and more consistent with $\epsilon=0$ for all values of $n$.

The results from constraining the evolution of the combined XLF data--sets
show a degeneracy in the opposite sense: higher values of $n$ demand higher
values of $\epsilon$. This relation can be clearly explained from the $\Omega=1$
plot alone. Again, the model scales the XLF in such a way that for $\epsilon=0$,
the result is an over--prediction of the cluster luminosities. However, this effect 
becomes less pronounced as $n$ {\it decreases}, leading to lower values of
$\epsilon$. Eventually, the local luminosity function is shifted so much that the
curvature becomes important, leading to an under--prediction of the abundance of
clusters at the faint--end, where most of the data--points lie. This effect places
a lower limit on the value of $n$. For lower values of $\Omega_0$, the data--points 
are pushed to brighter luminosities, as well as lower values of $dn/dL_{\rm x}$. 
Although the faint--end
data are consistent with no evolution for $\Omega=1$, the re--scaling forces the points
below the local XLF for lower values of $\Omega_0$. The brighter data (in particular
the \EMSS 2 point) is less sensitive to the change in cosmology, due to the steeper
slope of the XLF. Coupled with the fact that the evolution of structure is weaker, an
adequate fit to the data is allowed for more negative $n$.

In comparison to the open models, the flat ($\Omega_0+\Lambda_0=1$) models show slight
variations in the constrained region of parameter space. The reason for the
difference is two--fold. Firstly, the data--points themselves are scaled 
differently between cosmologies than their open counterparts. This is because
the distance to a fixed redshift is larger for flat models (hence also increasing
the volume element). This has the effect akin to having an open model with a 
lower value of $\Omega_0$. Secondly, the growth of structure in a flat model is
more rapid than the corresponding open case, which affects the model scaling
relations not unlike an open model with a higher effective $\Omega_0$.

Combining the constraints from the $L_{\rm x}-T$ and XLF measurements, we are unable
to determine the value of $\Omega_0$ (with or without $\Lambda_0$) from the data
alone: there is always a region of parameter space consistent with both observations. 
The same conclusion applies to all of the individual XLF data--sets. Specifically, the 
faint--end data--points from the \ROSAT surveys are unable to set tight constraints; 
however they are able to set a lower limit on the slope of the power spectrum that is 
sensitive to the value of $\Omega_0$. The \EMSS samples, particularly the \EMSS 2 
cumulative point, start to provide tighter limits on $n$ and $\epsilon$, since these 
results have started to probe the more sensitive luminosity range, brighter than 
$L_{\rm x}^*$. Therefore, future surveys will be more efficient at extracting information 
on the evolution of clusters by covering larger areas of sky, probing luminosities around 
and beyond $L_{\rm x}^*$. 

\subsection{Breaking the n--$\Omega_0$ degeneracy}
\label{sec:degen}

It is evident that the model is unable to determine a value of $\Omega_0$ from the
XLF and $L_{\rm x}-T$ normalisation constraints alone -- a feasible range of $n$ and
$\epsilon$ can always be found for the particular scenario. Noting that the constrained
range of $n$ is sensitive to $\Omega_0$, combining an independent determination of $n$
with the X-ray data will be able to place a significant constraint on the value of 
$\Omega_0$. There are two ways in which this can be achieved: a theoretical prediction 
of the primordial power spectrum that gives the (cosmology--dependent) slope on cluster 
scales or a direct measurement from redshift surveys. 

For the theoretical case, we need 
to obtain the local slope of the power spectrum on cluster scales, $n_{\rm eff}$ such that 
\begin{equation}
n_{\rm eff} = \left. {d {\rm ln} P \over d {\rm ln} k} \right| _{k_{\rm eff}}.
\label{eqn:pslope}
\end{equation}
Although there has been feverous debate recently, the currently most popular cosmogony
is that dominated by cold dark matter (CDM), with $n\rightarrow1$ for small $k$, turning
over to $n\sim-3$ on small scales. We have adopted the transfer function given by 
Bardeen et al. (\shortcite{BBKS}). The turnover scale is determined by the shape parameter,
$\Gamma$, which is in turn, determined by the cosmological model:
\begin{equation}
\Gamma = \Omega_0 \mbox{h} \, \exp \left[- \frac{\Omega_b}{\Omega_0} \, 
\left( \Omega_0+\sqrt{2 \mbox{h}} \right) \right].
\end{equation}
(\cite{Sugi95}). We take the value $\Omega_b=0.013 \, \mbox{h}^{-2}$ from nucleosynthesis constraints
(\cite{CST}). No contribution enters from the cosmological constant, due to the fact
that it contributed negligibly to the energy density at the last scattering epoch. To
estimate $k_{\rm eff}$, we assume that the virial mass of a typical rich cluster in the local
Universe (\eg the Coma cluster) is $M \sim 5\times10^{14}\hmsun$. If the cluster formed
by the top--hat collapse of a spherically overdense region, the comoving radius required to
contain this mass (assuming present background density) would be $R \sim 8\hMpc$ (\cite{ECF}).
This is only true for a critical density Universe but can easily be modified for lower densities,
since $R \propto \Omega_0^{-\frac{1}{3}}$. The value of $R$ can then be converted to an effective 
scale on the linear power spectrum by using the following formula (\cite{PD94}):
\begin{equation}
k_{R} = \left[ { \left[ \frac{1}{2} (n+1) \right] ! \, \over 2 } \right]^{1 \over n+3}
\, {\sqrt{5} \over R},
\end{equation}
which should iteratively converge on the values $(n_{\rm eff},k_{\rm eff})$, with the use of
equation \ref{eqn:pslope}. The calculated values of $n$ are represented as solid lines in 
Figure \ref{fig:megaplot}. Note that the value of $n$ becomes more negative with decreasing
values of $\Omega_0$. Although the effective scale of galaxy clusters is larger in 
a low mass--density Universe, the change in shape of the CDM power spectrum itself (parametrised
by $\Gamma$) dominates the shift in the value of $n$. As $\Omega_0$ decreases, the position
of the turnover moves to larger scales, leading to a more negative slope on cluster scales.
Figure \ref{fig:megaplot} shows the result of this calculation, plotted as a solid 
vertical line. 
Analysing the likelihood distributions for a continuous range
of $\Omega_0$, we find that the CDM prediction is only consistent with the X--ray data for 
$\Omega_0 < 0.55$ ($\Lambda_0=0$) and $0.1<\Omega_0<0.7$ ($\Omega_0+\Lambda_0=1$), 
to $95\%$ 
confidence. We can place a lower limit on $\Omega_0$ in the flat model because the
predicted value moves to more negative $n$ faster than the XLF contours, as $\Omega_0$
decreases. 

Another approach is to constrain $n$ directly by measuring the power spectrum on rich cluster
scales from wide--area sky surveys. One example (\cite{TED97}) comes from the APM cluster
sample, which has a median redshift  $\left< z \right>=0.09$, measuring a slope, 
$n_{\rm APM} = -1.6 \pm 0.3$. We currently assume that the underlying
power spectrum has negligible curvature on the range of scales we are studying. Ideally,
we would separate the population into redshift bins and scale each respective population relative
to the preceeding one, using the slope of the spectrum at that epoch. Imminent surveys such as the 
Sloan Digital Sky Survey and the 2dF Galaxy Redshift Survey should provide the necessary data to 
constrain the slope at higher redshifts. Furthermore, the APM result assumes that the bias is 
linear and therefore scale--invariant on scales of the mean inter--cluster separation. 

\begin{table}
\begin{tabular}{l|l|l|l|l}
\hline
Survey		& \multicolumn{2}{c}{$\Omega_{0,u} (\Lambda_0=0)$}
		& \multicolumn{2}{c}{$\Omega_{0,u} (\Lambda_0=1-\Omega_0)$}\\
\hline
\SHARC		& $1.0$ & $[0.6,>1.0]$	& $1.0$ & $[0.5,>1.0]$\\
\WARPS		& $1.0$ & $[0.75,>1.0]$	& $1.0$ & $[0.7,>1.0]$\\
\RDCS		& $1.0$ & $[0.65,>1.0]$	& $1.0$ & $[0.6,>1.0]$\\
\EMSS 1		& $1.0$ & $[0.7,>1.0]$	& $1.0$ & $[0.6,>1.0]$\\
\EMSS 2 	& $0.7$ & $[0.4,>1.0]$	& $0.65$& $[0.3,>1.0]$\\
\hline
COMBINED DATA	& $0.4$ & $[0.25,0.6]$	& $0.4$ & $[0.2,0.65]$\\
\hline
\end{tabular}
\caption{Upper limits on the value of $\Omega_0$ (with and without $\Lambda_0$), using the constraint
         that $n=-1.6$, from the APM survey (Tadros, Efstathiou \& Dalton, 1997). Results from
	both individual and combined XLF data--sets are presented. Limits are set using the $95\%$ 
         confidence limits consistent with both the XLF and $L_{\rm x}-T$ constraints.
         Ranges in square brackets are for quoted uncertainties in $n$ ($\pm 0.3$)}
\label{tb:om_n_upper}
\end{table}

To constrain the range of $\Omega_0$, we took the $95\%$ limits placed on $n$ and $\epsilon$
from the evolution of the XLF and $L_{\rm x}-T$ normalisation and subsequently fixed the value of
$n$ to $-1.6$. For the XLF data, we found that the model
fits the data more and more adequately as $\Omega_0$ decreases (for the range of $n$ 
considered): this is shown in the plot by the shift in the contours. Hence, this prevents
us from placing a lower limit on $\Omega_0$. In the flat models the geometry weakly 
alters the shape of the likelihood distribution but the overall trend is the same. Since
both constraints produce contours that become more aligned in lower $\Omega_0$ cosmologies,
we are unable to make further conclusions regarding the lower limit. 

What the data does allow us to do is place an upper limit on the value of $\Omega_0$.
The results are given in Table \ref{tb:om_n_upper}, displaying the values that 
are consistent with the $95\%$ confidence limits of the XLF and $L_{\rm x}-T$ relations.
Results for both the individual and combined XLF data--sets
are presented.  The ranges in brackets were determined by varying the value of $n$
by $\pm 0.3$. It is evident that the
value of $\Omega_0$ is unconstrained using the APM value, for all of the individual data-sets
except the \EMSS 2 survey.
The flat models give similar results to their open counterparts, with  
the range in the upper limits being slightly larger. Combining the XLF surveys sets a 
tighter constraint: $\Omega_0<0.6$ (or equivalently 
$\Omega_0<0.65$ for $\Omega_0+\Lambda_0=1$).

\subsection{Evolution of Cluster Core Sizes}

\begin{figure}
\centering
\centerline{\psfig{file=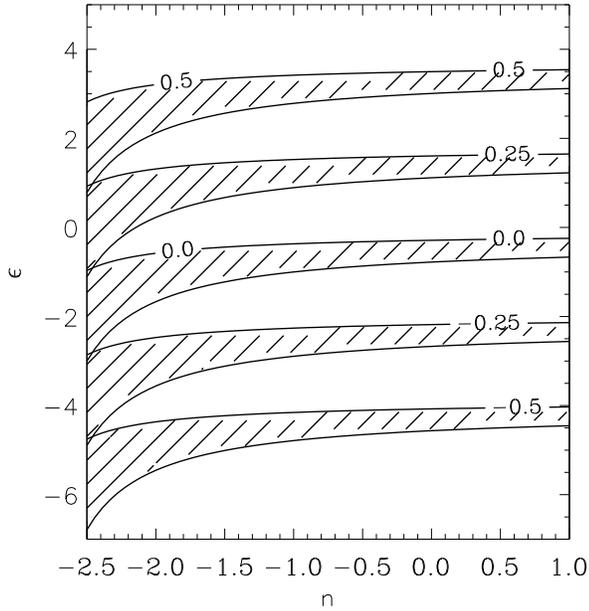,height=9cm}}
\caption{The region of parameter space covered by evolution of the cluster core radius
for various models. The shaded regions depict the range of cosmologies, from $\Omega=1$
(with the greatest amount of structure evolution) to $\Omega=0.1,\Lambda_0=0.0$ (with the least
amount of structure evolution). Each particular area illustrates the different amount of
evolution in the redshift range $z=[0.0,0.5]$, with the labels indicating the value
$\Delta{\rm log}r_c.$}
\label{fig:core}
\end{figure}

Finally, we consider another possible constraint that could be readily measured from
X--ray data: the evolution of cluster core--radii. Using the model we developed in
section \ref{sec:modelx}, we predict the characteristic core-radius of a cluster population
to scale as:

\begin{eqnarray}
r_c  &\propto&  \delta_+^{a} \, (1+z)^{b} \, (1+z_f)^{c} \, 
\left({\Delta_{vir} \over \Omega}\right)^{d},
\\ \nonumber
a&=&{2 \over n+3} \left( 1 - \frac{1}{\beta} \right), \\ \nonumber
b&=&{\epsilon \over 2\beta}, \\ \nonumber
c&=&{1 \over 2\beta}-1, \\ \nonumber
d&=&\frac{1}{3} \left( {1 \over 2\beta} -1 \right).
\label{eqn:core}
\end{eqnarray}
 
Even though the core radius depends on both $n$ and $\epsilon$, it's evolution
is determined much more sensitively by the changes in the central entropy of the gas rather
than the amount of structural evolution. Figure \ref{fig:core} illustrates the regions
of parameter space consistent with various amounts of core radius evolution, 
out to a redshift $z=0.5$. The 
background cosmology and spectral index have only a weak influence on the evolution of $r_c$, 
hence observations should quite easily place a constraint on the value of $\epsilon$ -- the two
extremes of the shaded regions represent scaling in the core by a factor of $\sim 0.3$ and $3$
respectively. Core sizes that shrink with redshift demand negative values of $\epsilon$ 
(\ie the core entropy of clusters decreases with redshift) whereas core sizes that grow
with redshift demand values of $\epsilon$ that are weakly negative or higher. 
This approach is useful in the sense that we can directly measure the
rate of entropy evolution in the core and test it's consistency with the other methods
presented above. 

Analysis involving looking at changes in the cluster core size has now started 
to appear in the literature. Results are currently suggesting no significant evolution in the 
core size, at least out to $z \sim 0.5$ (\cite{Vik98}). Given the uncertainties, our results 
are consistent with this, however positive evolution in the cluster core radii ($\epsilon>0$) 
would be preferred. At present, the sample sizes are small: a more comprehensive measurement 
of core radius evolution will serve as an important test of the validity of this model.

\section{Discussion}
\label{sec:discuss}

\begin{figure*}
\centering
\centerline{\psfig{file=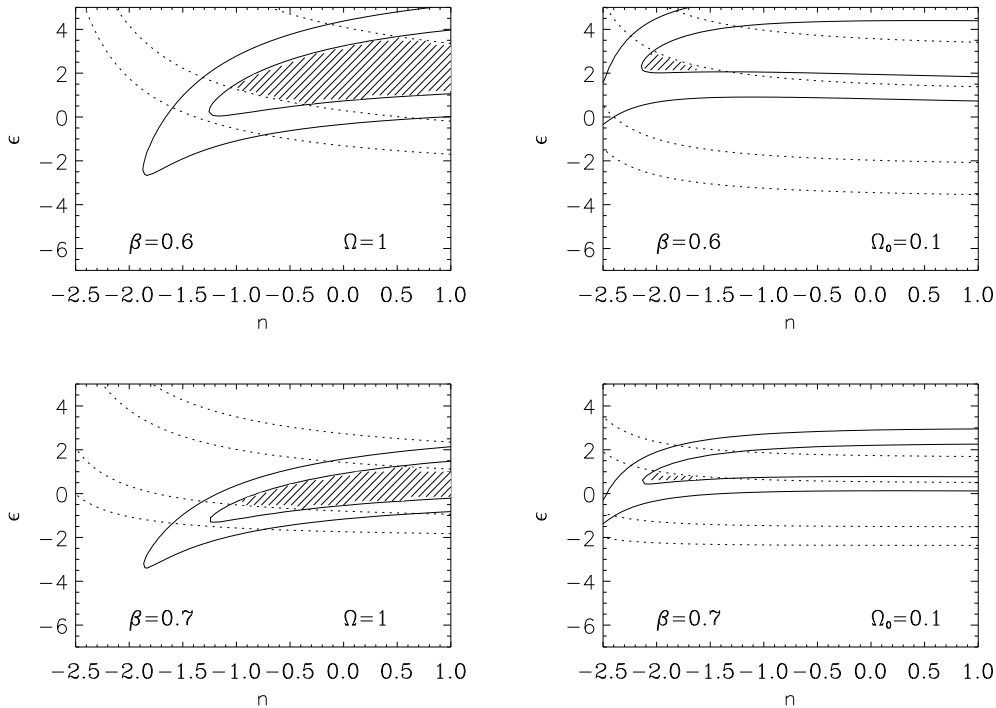,height=10cm,width=14cm}}
\caption{An example of the effect of varying $\beta$ using the \SHARC XLF data. The left
column is for $\Omega=1$ (the most mass evolution) and the right column $\Omega_0=0.1$
(the least mass evolution). The top and bottom rows for the expected limits on $\beta$
(0.6 and 0.7 respectively). The effect on $n$ is insignificant, although raising $\beta$
lowers the range of $\epsilon$.}
\label{fig:betaplot}
\end{figure*}

At this stage, we go on to focus on several issues that are worthy points for discussion.
Firstly, our results are based on several important assumptions made
when modelling the X--ray evolution of clusters, so we discuss our choices and
their robustness. Secondly, particularly due to the recent publication of \ROSAT and \ASCA
data, several results constraining $\Omega_0$ using the evolution of X--ray clusters 
have now appeared in the literature. We therefore compare our results with those found
by other groups. 
Finally, we address what we think is an important strategy for observations (particularly with
the imminent launch of the \XMM satellite, and pin--point the areas that we think will maximise
future success in this field.

\subsection{Model Assumptions}
Although general and based on physically motivated scaling
relations, there are several key approximations in the model used to 
arrive at this conclusion. 
One example is the $\beta$ profile: the results presented in \S3
assume that the X--ray emission
profiles are well fit when using the $\betamodel$, where the density of the 
intracluster gas tends to a constant value at small r (the core) and the 
isothermal result at large r, for a value of $2/3$. 
Individual measurements of $\beta$ do not necessarily give this value, although we assume
that the average value over the population at a given epoch is well represented by 
$\left< \beta \right> =2/3$. However, it is important to investigate the
tolerance of the model when varying the value of $\beta$. Figure \ref{fig:betaplot} illustrates the
effect on the constrained range of $n$ and $\epsilon$ that variations in $\beta$ induces.
Using the \SHARC data as an example, we have taken two values of $\beta : 0.6$ and 
$0.7$, defining what we think are the expected limits of any
fluctuation. Two cosmological models have been chosen : $\Omega=1$ and $\Omega_0=0.1$, which
represent the most and least amount of structural evolution respectively.  Changing $\beta$ 
can significantly change the range of $\epsilon$ values but
has relatively little effect on the values of $n$. This, coupled with the work on the core radius
discussed in the last section, reinforces the idea that both $n$ and $\epsilon$ contribute
separately to the evolution of clusters: the entropy evolution is sensitive to the way
in which the gas is distributed, with $\beta$ and $r_c$ defining the shape of the gas density 
profile, while the mass evolution is tied in with the background density. Hence, the dependence
of $\Omega_0$ on $\beta$ is weak, although certain models may be ruled out on the basis that
the range of $\epsilon$ may be implausible. This is evident in the figure, where 
the combination $(\beta=0.6,\Omega_0=0.1)$ pushes allowed values of $\epsilon$ into the
limit set by the maximum cooling calculation ($\epsilon=2.0$) given in Paper~I. 

With the Launch of the \AXAF satellite, it will become possible to study
the emission profiles of distant clusters in more detail. These
studies may show that the cluster profile evolves with redshift.
For example, simulations of cluster evolution with pre--heated gas by Mohr \& Evrard (\shortcite{Mohr97})
suggest that the entropy injected reduces the overall slope of the
gas density profile rather than just that of the profile of the very centre.
This effect can be parameterised by a much slower role over near the core
radius but the scaling of the solution remains the same, since the 
X--ray emission still scales as $\rho_c^2 r_c^3 T^\alpha$.
Thus, as Figure~\ref{fig:betaplot} shows, the $\beta$ profile
and the $\epsilon$ parameter are interconnected -- while observations
of the emission profiles of high redshift clusters would cause
us to revise our physical interpretation of a particular $\epsilon$
value, and re--examine the balance of heating and cooling within the
cluster, it would not invalidate the phenomenological description
provided by the model.  

We may apply the same argument to take into account the non--isothermal 
temperature of structure of clusters suggested by
\cite{Marketal98}. The model only requires that the temperature
structure is normalised by the virial temperature of the dark matter
halo. The temperature structure may affect the rate at which the
cluster core evolves for a given entropy change, but this should again taken
into account when the $\epsilon$ parameter is interpreted.

The model has been developed under the assumption that
clusters of galaxies are in dynamical equilibrium. Clusters
that are seen during major mergers may have X--ray luminosities and average
temperatures that are far from their hydrostatic values. Nevertheless,
the description we have developed applies to the cluster population
as a whole, and the properties of individual clusters may differ wildly
from the average. In this
sense, the effects of mergers are already incorporated through the entropy 
evolution described by the model, and the disruption of hydrostatic
equilibrium is mimicked by an increase in core entropy. The cluster 
temperatures are
also susceptible to these departures, however, since they are weighted
by the luminosity. Ideally, we should use cluster temperatures measured
from the outer regions of clusters where the gas temperature better
reflects the virial temperature of the gravitational potential. These
are not available for distant clusters at present, but our results will
still be valid so long as the evolution of the luminosity weighted temperature
reflects the evolution of the cluster potential. 
This is a problem common to all methods based on the X--ray temperatures of clusters.

The slope of the luminosity--temperature relation does not explicitly
enter our determination for the evolution of clusters. All that is required
in our model is the evolution of the normalisation. We have
implicitly assumed that the slope remains fixed at its present--day value. 
This assertion, which corresponds to the assumption that the entropy
evolution parameter ($\epsilon$) is independent of cluster temperature,
is unavoidable with the limited temperature data that is available for high
redshift clusters. Consequently, it is important that the evolution of 
the normalisation is determined by using consistent slopes for high and low
redshift data, and preferably using high redshift clusters representative
of the clusters used to determine the luminosity function evolution.
At present, temperature measurements are available only for the brightest
clusters at high redshift. This situation will hopefully improve with
the launch of the \XMM satellite (\cite{Lumb96}); its large collecting 
area is ideally matched to the determination of temperatures for the lower 
luminosity, distant clusters.
The slope of nearby ($z<0.05$) clusters has been most 
recently determined by Markevitch (\shortcite{Mark98}), using \ROSAT luminosities and 
\ASCA temperatures, measuring a value, $\lambda \sim 2.65$. However, we make use of the older sample
based on \Einstein MPC data (\cite{David93}) for two reasons. Firstly, constraints on
the evolution of the $L_{\rm x}-T$ relation are provided by Mushotzky \& Scharf (\shortcite{MS97}), 
which uses this (local)
sample to contrast with their higher redshift 
($\left< z \right> \sim0.3$) \ASCA data. Also, the slope supplied by
Markevitch has been calculated for clusters with cooling flows removed, in order to
directly compare with models that exclude non--radiative components in the plasma
evolution. Since our model is based on evolution of the central entropy, we cannot
exclude the cooling flow clusters, which probably contribute to most, if not all
of the negative entropy evolution in cluster populations. We argue that the dominant cause
for the intrinsic scatter in the relation is due to each cluster having it's own particular
core entropy, coming from their individual formation histories.  

The considerations we discussed above related to the distribution
of gas within the dark matter potential. In a low density universe,
it is necessary to explicitly account for the distinction between
the background density of the Universe when the cluster is formed,
and it's value when the cluster is observed. We have related these
two epochs by assuming that the mass of the cluster grows by a 
factor of two (Kitayama \& Suto, 1996). However, for $\Omega_0 > 0.1$,
this factor produces only a small change in the halo evolution.

\subsection{Comparison with Other Methods}
A number of other papers have recently appeared in the literature
dealing with the constraints on $\Omega_0$ from the X--ray evolution of clusters.
One of the most popular methods uses the evolution of the 
cluster abundances (\eg \cite{HENRY}, \cite{ECFH}, \cite{VL}, \cite{Math98}, \cite{Ouk97}, \cite{Reich}).
To extract the value of $\Omega_0$, an estimator for the cluster virial mass
is required. This can be achieved by calculating the cluster temperature 
function at any given epoch and converting this to mass using the 
Press--Schechter formalism with the assumption that the cluster haloes form 
via spherical infall. The result from this method is
still slightly unclear -- at present the majority of authors find values of $\Omega_0$ 
typically in the region $0.3-0.5$ with or without $\Lambda$, although some 
authors find higher values including $\Omega=1$ (\eg \cite{S98}, \cite{BB98}). 
An extensive discussion on the existence of this clash in 
results can be found in Eke et al. (\shortcite{ECFH}). In particular, they
highlight the discrepancy in the abundance of low--redshift clusters, particularly
the high temperature end, where the measured value is uncertain. Those favouring
high $\Omega_0$ assume a higher abundance of clusters in this region than those
favouring low values of $\Omega_0$. 

Our work differs from these treatments in that it does not assume a 
detailed model for the abundance of clusters or for their temperature
or X--ray luminosity distribution. Instead, we have emphasised the
scaling relations that must relate the properties of clusters at one
epoch with those at another. By introducing an additional
parameter that describes the evolution of the core gas entropy,
our approach allows us to inter--compare clusters at different epochs with
minimal further assumptions. Comparing our results with those obtained by other authors
illuminates the constraints which are general and those which are
model--specific. As an example, Reichart et al. (\shortcite{Reich}) also look at
luminosity function evolution by adopting the Press--Schechter approach for the
evolution of the mass function , but use the $L_{\rm x}-T$ relation to calibrate 
their mass to X--ray luminosity relation. In addition, their model 
differs from ours in that it involves the slope of the $L_{\rm x}-T$ relation 
explicitly, resulting in a purely empirical calibration of the luminosity--mass 
conversion. By contrast, our approach uses the weak self--similarity 
principle (plus the entropy evolution) to scale the properties of clusters
between epochs. As discussed extensively in Paper~I, this is completely 
separate from the scaling relations between clusters of different mass at the same 
epoch. Trying to equate the two requires that we adopt a Strong 
Self--Similarity principle; something for which there is little physical
justification. Reichart et al. adopt the same parameterization as Evrard \& Henry
(\shortcite{EH91}), notably $L_{\rm x} \propto T^{p} (1+z)^{s}$; assuming $\beta = 
2/3$, we find for our model that $p=13/6$ and $s=(13-3\epsilon)/4$.
However, as we have emphasised, the way in which we constrain our parameters (particularly
$\epsilon$ in this context) is fundamentally different, and we should not expect to obtain
identical results. 

We choose to concentrate our comparison with the work of Eke et al. (\shortcite{ECFH}).
These authors derive a value of $\Omega_0$ from the evolution of the cluster
temperature function, using emission--weighted temperatures obtained from \ASCA 
data (\cite{HENRY}). Since the cluster sample is X--ray flux limited, they have
to correct for incompleteness in their estimation of the temperature function.
This is done by assuming a non--evolving $L_{\rm x}-T$ relation (\ie fixed slope,
normalisation and scatter). The advantage of this method is that it infers the
mass distribution in a more direct manner than our approach, hence their work 
places a strong constraint on $\Omega_0$: $\Omega_0 = 0.43\pm0.25$ for $\Lambda_0=0$ 
and  $\Omega_0 = 0.36\pm0.25$ with a non--zero cosmological constant. This was
achieved with a substantially smaller data--set. Furthermore, the authors were
able to internally constrain the shape of the power spectrum. This is possible
because they directly link the form of the temperature function to the mass function 
through the Press--Schechter model (assuming a CDM power spectrum).
They arrive at a value of $\Gamma$ of $0.08 \pm 0.07$ ($\Gamma=0.09 \pm 0.08$ for
$\Omega_0 + \Lambda_0 =1$).

To compare our results to those of Eke et al., we first convert their measured range
of $\Gamma$ to an equivalent range of $n$, using the same method discussed in
section \ref{sec:degen}. We then find the range of $\Omega_0$ that is both consistent
with the X--ray data and the calculated range of $n$. We find that $\Omega_0 < 0.45 
(\Omega_0<0.5$ for $\Omega_0+\Lambda_0=1$) to $95\%$ confidence, in agreement with their 
upper limit. 
However, we fail to place a constraint on the lower limit of $\Omega_0$. This is initially 
surprising, but seems to be due to systematic differences in the data.
We find that, for a low $\Omega_0$ cosmology, the available data
are consistent with weak negative evolution of the XLF and no evolution in the 
$L_{\rm}-T$ normalisation.
By contrast, Eke et al.'s temperature function data (corrected for
cosmology) shows a decline in temperature function amplitude suggesting 
that a very low value of $\Omega_0$ is unacceptable due to the deficit of clusters at this
epoch. Such dependence on the particular data set is worrying, but emphasises the importance 
of using independent data--sets to establish consistent conclusions and reduce
systematic biases. It is satisfying to find that our method finds similar results, at least
in determining the upper limit.

\subsection{Future Directions}
\label{subsec:future}

Since the results presented in this paper were based on scaling the \ROSAT BCS XLF 
to fit the higher redshift surveys, we have assumed that the two populations are well 
separated in redshift. This is not immediately obvious, since the \ROSAT BCS sample 
has clusters out to $z \sim 0.3$, where the other samples start around this value. 
The saving grace is that the higher--redshift surveys have detected lower luminosity 
clusters ($L_{\rm x} < L_{\rm x}^*$), due to the small solid angles of the surveys,  
whereas the BCS clusters in this regime are at much lower redshifts. However, this range 
of luminosities places a much weaker constraint on the model than if high redshift data 
were available at the brighter end ($L_{\rm x} > L_{\rm x}^*$). Improved coverage will 
result in a better internal constraint, since the effects of evolving cluster number density 
and cluster luminosity will be more easily separated. As a pointer for the future, what 
is required is a more statistically sound description of the high--redshift XLF from 
$L_{\rm x}^*$ and above. As an hypothetical example, we assume that the combined data--set
used in this paper is a good description of the XLF for luminosities around and
below $L_{\rm x}^*$. A measurement is then made at  
$L_{\rm x} (0.5-2.0 \keV) \sim 10^{45} \ergs$, detecting 10 clusters at 
$\left< z \right> = 0.5$ in a solid angle that resulted in a cumulative abundance
in agreement with the BCS cumulative XLF ($N(>L_{\rm x}) \sim 0.25 \times 10^{-8} \Mpc^{-3}$).
This strongly suggests a detection of no evolution, particularly for the
$\Omega=1$ geometry.
Using the APM constraint on $n$, we find that $\Omega_0<0.2$ at the $99\%$ level. 
In practice, the \XMM satellite is particularly 
suited to this task, with its large collecting area and high spatial resolution, 
that will allow one to easily make the distinction between clusters and AGN. 
What is required, however, is a wide area survey covering at least 500 square degrees, 
rather than a survey that is particularly deep.  In the near term, progress will be made 
through the \ROSAT North Ecliptic Pole survey -- a wide angle survey composed of superposed 
\ROSAT All--Sky Survey strips (\cite{HENRYNEP99}).

\section{Conclusions}

In this paper, we have taken the entropy--driven model of cluster evolution, detailed
in Bower (1997) for an $\Omega=1$ Universe, and modified it to account for evolution
in different cosmological scenarios, specifically open models and flat, non--zero
$\Lambda$ models. This allows us to separate contributions made by the hierarchical
growth of structure (controlled by the slope of the power spectrum, $n$) and changes
in the core entropy of the intracluster gas (controlled by the entropy evolution 
parameter, $\epsilon$).

We then placed constraints on $n$ and $\epsilon$ for seven reasonable cosmological
models, using the current wave of X--ray data for the two observables:
the X--ray Luminosity Function and the Luminosity--Temperature relation. For the 
$L_{\rm x}-T$ relation, we have taken the compiled low and high redshift
samples from David et al. (1993) and Mushotzky \& Scharf (1997) respectively and
determined the best--fit slope and scatter. We subsequently placed 68\% and 95\% confidence
limits on the evolution of the temperature normalisation, which is slope--dependent.
For the XLF, we have used the recently available \ROSAT samples, specifically the SHARC
(\cite{Burke97}), WARPS (\cite{J98}) and RDCS (\cite{Ros98}) high--redshift, 
non--parametric determinations,
as well as the older \EMSS samples (\cite{H92}, \cite{LG95}). The amount of evolution in these
measurements was quantified by comparing with the local \ROSAT BCS XLF (\cite{E97}). Using
the X--ray data alone, we find acceptable regions of parameter space for every 
cosmological model considered, to a high degree of confidence. 
Specifically, as the density parameter decreases, more negative values of $n$ are allowed
to compensate for the weakening growth rate of clusters. Values of constrained $\epsilon$ 
are appropriate to modify the luminosity evolution to compensate for the measured change 
in abundance, moving weakly from positive values to zero (no evolution) for low--density Universes. 

To break the degeneracy between $n$ and $\Omega_0$, we calculated
the slope of the power spectrum based on the CDM paradigm, by taking the typical mass of
a rich cluster and calculating the corresponding fluctuation scales that give rise to such
objects. We find that the CDM calculation sets a limit of $\Omega_0 < 0.55$ 
($0.1<\Omega_0 <0.7$ for $\Omega_0+\Lambda_0=1$). Using the value determined by the APM
survey ($n=-1.6\pm 0.3$), we conclude that $\Omega_0 < 0.6 \ (<0.65$ for 
$\Omega_0+\Lambda_0=1$). All limits were calculated to $95\%$ confidence. In general,
we cannot 
determine a lower limit to the value of $\Omega_0$, since our model fits the data much more 
comfortably at fixed $n$, for lower density Universes.
The evolution of the cluster core--radius was also considered, where we showed that
it depends much more sensitively on the entropy evolution of clusters than
the structural evolution (determined by $n$ and $\Omega_0$). Present
data suggest no strong evolution out to $z\sim 0.5$ (\cite{Vik98}), roughly
consistent with our results given the quoted uncertainties.

Finally, we discussed the robustness of assumptions we have made in the model,
particularly the effect of varying the mass distribution of the gas
(controlled by the value of $\beta$). Again, this has a much stronger effect
on the constrained range of $\epsilon$ than the rate of structural evolution, and hence
only weakly affects the determination of $\Omega_0$. We then compared our results to
that of other groups, particularly with Eke et al. (\shortcite{ECFH}). We find that 
our upper limit is consistent ($\Omega_0<0.45$, or $\Omega_0<0.5$ for $\Omega_0+\Lambda_0=1$
at $95\%$ confidence) although we cannot place a lower limit on $\Omega_0$.

In order to provide tighter limits on the value of $\Omega_0$, it will be essential
to obtain an accurate measurement of the bright end of the XLF out to redshifts at least
comparable to present distant cluster samples, ideally from a wide angle survey. 
This is something we hope that the forthcoming \XMM mission can provide.

\section*{Acknowledgements}
The paper would not have been completed without the generous help of Doug Burke,
Piero Rosati and Laurence Jones with their XLF data and Vince Eke, Pat Henry, 
Shaun Cole and Carlos Frenk for useful discussions.
This project was carried out using the computing facilities supplied by the
Starlink Project.
STK and RGB acknowledges the support of a PPARC postgraduate studentship and
the PPARC rolling grant for ``Extragalactic Astronomy and Cosmology at Durham''
respectively.

\end{document}